# Metal-insulator-metal transition in NdNiO$_3$ films capped by CoFe$_2$O$_4$

M. S. Saleem, C. Song,[*] J. J. Peng, B. Cui, F. Li, Y. D. Gu, and F. Pan[†]

Key Laboratory of Advanced Materials (MOE), School of Materials Science and Engineering, Tsinghua University, Beijing 100084, China

Metal-insulator transition features as a transformation, from a highly charge conductive state to another state where charge conductivity is greatly suppressed when decreasing the temperature. Here we demonstrate two consecutive transitions in NdNiO$_3$ films with CoFe$_2$O$_4$ capping, in which the metal-insulator transition occurs at ~85 K, followed by an unprecedented insulator-metal transition below 40 K. The emerging insulator-metal transition associated with a weak antiferromagnetic behavior is observed in 20 unit cell-thick NdNiO$_3$ with more than 5 unit cell CoFe$_2$O$_4$ capping. Differently, the NdNiO$_3$ films with thinner CoFe$_2$O$_4$ capping only exhibit metal-insulator transition at ~85 K, accompanied by a strong antiferromagnetic state below 40 K. Charge transfer from Co to Ni, instead of from Fe to Ni, formulates the ferromagnetic interaction between Ni–Ni and Ni–Co atoms, thus suppressing the antiferromagnetic feature and producing metallic conductive behavior. Furthermore, a phase diagram for the metal-insulator-metal transition in this system is drawn.

[*] songcheng@tsinghua.edu.cn
[†] panf@mail.tsinghua.edu.cn

The metal-insulator transition (MIT) is one of central subjects in the physics of correlated electron phenomena and transition-metal oxides. In regard of MIT, rare-earth nickelate[1] (RNiO$_3$, here R is rare earth element), SrVO$_3$[2] and VO$_2$[3] are typical materials in oxides family, offering the opportunity to explore true underlying physics and the feasibility of creating electronic devices such as resistive switches and sensors. The MIT in RNiO$_3$ has drawn increasing attention recently, not only due to their fundamental electronic structure like electrons interaction between nickel 3$d$ and oxygen 2$p$ bands, but also driven by the promise to emulate cuprate superconductors.[1-4] The electronic behavior of RNiO$_3$ has been explained by many researchers under Mott-Hubbard type[5,6] or charge-transfer type materials.[1] The room-temperature metallic state in RNiO$_3$ arises due to the overlapping of O-2$p$ valance band and Ni-3$d$ conduction band,[7,8] and at low temperature they usually behave as an insulator (except LaNiO$_3$).[9] Some published reports revealed that the low temperature insulating phase was suppressed completely under high-pressure-resistivity measurement, which is ascribed to an increase of electronic bandwidth and decrease of the charge-transfer gap $E_g$.[10,11] Other important factors such as, charge ordering associated with structural symmetry, and strain induced orbital polarization can also account for MIT.[12,13] Recently it is found that the capping of ultrathin LaAlO$_3$ enhanced the conductivity of ultrathin LaNiO$_3$,[8] reflecting the critical role of the capping effect.

In fact, the coupling of several degrees of freedom (such as spin, charge, lattice, and orbital) enriches the physical properties at the oxides interface, accompanied by the observation of numerous fascinating phenomena, such as charge transfer, orbital reconstruction and magnetic coupling etc.[12-17] Such charge transfer brings about exchange bias in paramagnetic-ferromagnetic bilayers and ferromagnetism induced in RNiO$_3$.[14–16] These investigations are generally carried out in perovskite-perovskite materials, e.g., La$_{0.75}$Sr$_{0.25}$MnO$_3$/LaNiO$_3$, CaMnO$_3$/LaNiO$_3$, LaMnO$_3$/LaNiO$_3$.[14–16] It is interesting to explore the interfacial effect of perovskite/spinel bilayers, a system extensively used for the nanostructures with vertical growth mode.[18,19] The experiments below demonstrate the unexpected metal-insulator-metal transition (MIMT) in perovskite NdNiO$_3$ (NNO) thin films with a spinel CoFe$_2$O$_4$ (CFO) capping layer, for which the transport can exactly reflect the behavior of the NNO with the highly insulating feature of the CFO taken into account.

NNO(20)/CFO($t$ = 0, 3, 5, 8, 15, and 18) (thickness in unit cell, u.c.) heterostructures were grown by pulsed laser deposition (PLD) from stoichiometric NdNiO$_3$ and CoFe$_2$O$_4$ targets. NNO ($\alpha_{NNO}$ = 3.81 Å) was grown epitaxially on LaAlO$_3$ ($\alpha_{LAO}$ = 3.79 Å) substrates at 770 °C under 100 mTorr oxygen pressure, and CFO ($\alpha_{CFO}$ = 8.391 Å) was deposited on the top of NNO at 660 °C under 30 mTorr oxygen pressure. LAO was chosen as the substrate because the compressive strained NNO on LAO is more conductive compared to the tensile strained films grown on SrTiO$_3$.[20, 21] The growth process was monitored by reflection high-energy electron diffraction (RHEED). Then the samples were cooling down under 500 Torr oxygen pressure with a speed of 10 K/minute. The morphology of the samples was characterized by atomic force microscope (AFM). Room-temperature $L$-edge x-ray absorption spectroscopy (XAS) for cobalt (Co), nickel (Ni) and iron (Fe) was performed in total electron yield mode (TEY) to analyze the charge transfer. For transport measurements, the films were then patterned into Hall-bar structure with 600 μm effective length and 50 μm width by optical lithography and Ar etching process. The longitudinal resistivity $\rho_{xx}$ in the range of 300–10 K was measured in a physical property measurement system (PPMS). The superconducting quantum interference device (SQUID) magnetometer was used to measure the magnetic properties of the samples with in-plane fields.

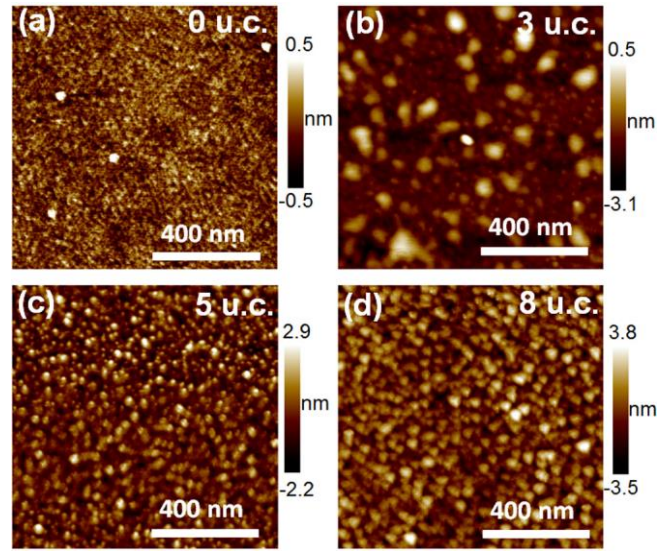

**FIG. 1** Images of surface morphology of NNO/CFO heterostructure, where (a) 0 u.c. represents only 20 unit cell NNO sample without CFO cap, while 3 (b), 5 (c), and 8 (d) u.c. shows the CFO thickness in term of unit cell on the top of 20 unit cell NNO.

Figure 1 shows the AFM surface morphology of NNO single layer and NNO/CFO bilayers. In Fig. 1(a), the surface of the NNO single layer (without CFO) is smooth with a root-mean-square roughness ($R_{rms}$) of ~0.14 nm, reflecting the high quality growth of the NNO films on LAO substrate. The sample with 3 u.c. thick CFO exhibits larger grains which are not homogeneously distributed and reflect an island growth of the ultrathin CFO (in Fig.1 (b)). Apparently the NNO surface is not fully covered by CFO in this case. This island growth is ascribed to the poor wettability of CFO, for which CFO atoms diffuse from surrounding and make a droplet like nucleation.[22] This poor wettability nature between perovskite/spinal materials makes it a typical system for the investigations on nano-pillars and their coupling effects at the vertical interface.[22,23] Probability of droplet like nucleation occurs only in ultrathin layer with slow cooling.[24] The situation differs dramatically when the CFO is up to 5 unit cell. For the 5 u.c. sample, the CFO layer just covers the full surface of NNO, while in the 8 u.c. sample CFO perfectly covers the whole surface of the NNO layer, accompanied by the uniform distribution of grains with $R_{rms}$ of ~1.06 nm. The samples with thicker CFO, e.g., 15 and 18 u.c., show a similar behavior.

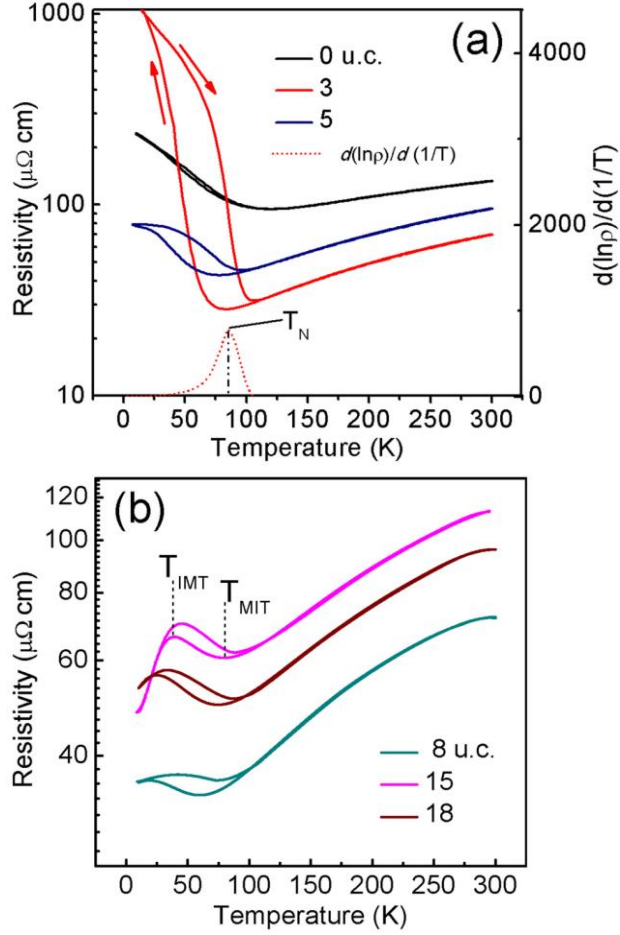

**FIG. 2** (a) and (b) are the $\rho$-$T$ curves of NNO films with different CFO capping layer thickness, 0 to 18 numbers represent the CFO unit cell at the top of 20 u.c. NNO. The dotted line shows the $d(\ln\rho)/d(1/T)$ versus T plot for 3 u.c. sample, derived from heating curve. The peak value of red dotted line indicates the Neel transition temperature ($T_N$). In Fig. 2 (b) $T_{MIT}$ and $T_{IMT}$ are two transition temperatures, metal-insulator and insulator-metal respectively. Upward and downward arrows indicate the cooling and heating sequence of $\rho$-T curves respectively.

The temperature dependent resistivity of the NNO/CFO heterostructures is presented in Fig. 2. For these measurements, the temperature decreases initially from 300 K to 10 K and then goes back to 300 K to acquire a loop of resistivity curve. As expected, all the NNO films with or without the CFO capping layer exhibit MIT at around $T_{MIT}$ = 85 K (metal-insulator transition temperature) with decreasing temperature. Apparently, the NNO films without CFO capping or only capped by 3 u.c. ultrathin CFO show the continuous increase of resistivity down to 10 K. Note that the resistivity of the NNO/CFO keeps almost constant at low temperature when the CFO is up to 5 nm. The situation turns out to be dramatically different when the CFO cap is thicker, e.g., 8, 15, and 18 u.c. After the MIT at 85 K, the resistivity decreases anomalously at lower temperature, showing an insulator-metal transition (IMT) with $T_{IMT}$ ~ 40 K (insulator-metal transition temperature). We then conclude that there appears MIMT in the 20 u.c.-thick NNO films when the CFO capping layer is thicker than 5 unit cells.

Hysteresis loops were measured to elucidate the influence of the CFO capping on the magnetization of NNO/CFO heterostructures. Considering the typical metal-insulator transition and metal-insulator-metal transition shown in the samples with 3 and 15 u.c-thick CFO respectively, the hysteresis loops for these two representative samples at different temperatures were measured and presented in Fig. 3(a) and (b), where the diamagnetic contribution of the LaAlO$_3$ substrate was subtracted from the raw data. The coupling between ferrimagnetic CFO and antiferromagnetic NNO generates exchange bias in the NNO/CFO system. For the 3 u.c. sample, the exchange bias field ($H_{EB}$) is greatly enhanced when the temperature decreases. The corresponding $H_{EB}$ are 0, 25, 150, and 270 Oe at 300, 90, 45, and 10 K, respectively. Obviously, the prominent $H_{EB}$ emerges at 45 and 10 K. In general, as the temperature goes down from 300 K to 10 K, the rare-earth nickelate undergoes a transition from a paramagnetic (PM) phase to an antiferromagnetic (AFM) phase at the Neel temperature $T_N$.[20,25]

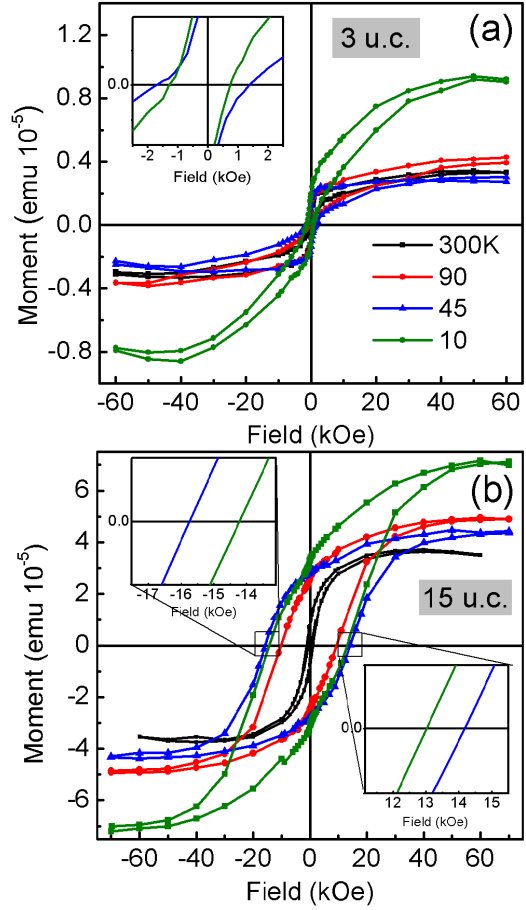

**FIG. 3** (a) and (b) represent respectively the hysteresis loops of 3 and 15 u.c. samples at different temperatures after 1 T cooling field. Inset is the large scale view of 45 and 10 K coercivities.

In the 3 u.c. sample, the Neel point is around 85 K, according to the peak value of the dotted line derived with $d(\ln\rho)/d(1/T)$ in Fig. 2(a), which is consistent to the observation of weak $H_{EB}$ at 90 K to some extent. The strong enhancement of $H_{EB}$ with decreasing the temperature from 90 to 10 K indicates that the antiferromagnetic behavior of NNO becomes more robust. Differently, in the 15 u.c. sample, the $H_{EB}$ is initially enhanced from 400 to 800 Oe with the temperature decreasing from 90 to 45 K [Fig. 3(b)], and then decreases to only 600 Oe with the temperature further decreasing to 10 K, as displayed in the inset of Fig. 3(b). The anomalous decrease of the exchange bias at 10 K reflects the weak antiferromagnetic behavior of NNO compared to its counterpart at 45 K. In addition, the dramatic increase of saturation magnetization at 10 K may also support the existence of the proximity effect between two layers at low temperature and the consequent decrease of the AFM nature of NNO.

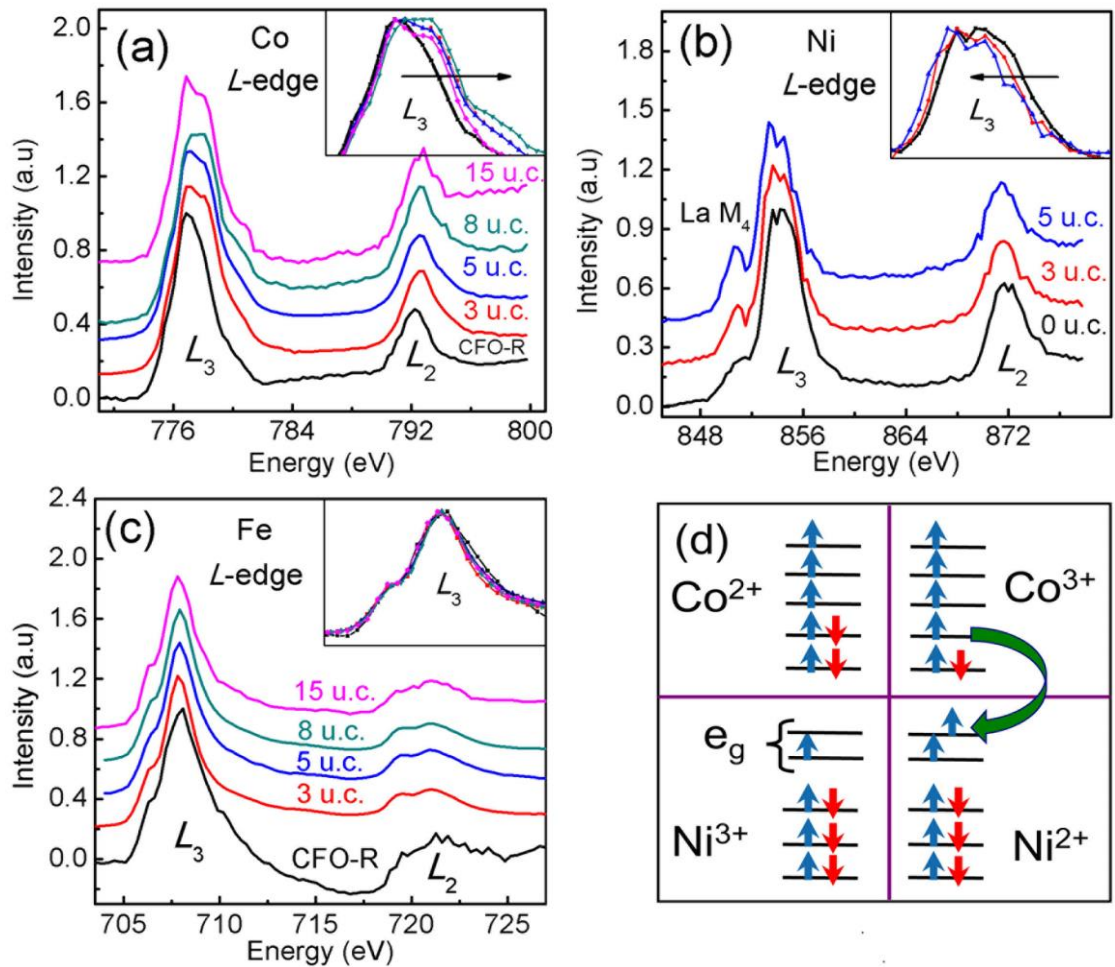

**FIG. 4** (a) (b) and (c) Normalized XAS spectra of Co, Ni and Fe respectively. The arrows in insets indicate the $L_3$-edge trend shift. (d) The proposed model of charge transfer ($Co^{2+} + Ni^{3+} \rightarrow Co^{3+} + Ni^{2+}$), the total spin moment of $Ni^{2+}$ increases due to high spin arrangement of electron in $e_g$ orbit.

To understand the evolution of the transport and magnetic behaviors, element-resolved x-ray absorption spectroscopy (XAS) in total electron yield (TEY) mode was used to characterize the electronic structure of NNO affected by the capping CFO. The inverse spinel $CoFe_2O_4$ has eight tetrahedral cation "A" sites occupied by $Fe^{3+}$ ions, while the 16 octahedral cation "B" sites are randomly occupied by eight $Co^{2+}$ and eight $Fe^{3+}$ ions. Fig. 4(a) displays the Co *L*-edge XAS of the NNO/CFO heterostructures, where the reference spectrum of the CFO single layer (CFO-R) grown on LAO substrates is also shown for a comparison. Remarkably, the *L*-edge peaks of the NNO/CFO ($t$ = 0, 3, 5, 8) heterostructures shift toward the higher energy direction in comparison to that of the CFO single layer, as highlighted in the inset of Fig. 4(a). Such a shift indicates that the valence state of Co ions in the bulk CFO is higher than the $Co^{2+}$ atoms in the vicinity of the NNO/CFO interface. This is further supported by the fact that the *L*-edge peak shifts toward the lower energy direction ($Co^{2+}$) when the thickness of the CFO layer increases from 8 u.c to 15 u.c. (~12 nm) where only Co ions suffering little interfacial influence were detected.[26] This is because the probing depth of TEY-XAS experiments is exponentially decaying within a few nanometers.[27]

Figure 4(b) is the Ni *L*-edge XAS spectra of the NNO reference sample and NNO/CFO heterostructures. The most eminent feature is that the Ni *L*-edge peak shift toward lower energy direction when the CFO thickness increases from 0 to 5 u.c., indicating the change of $Ni^{3+}$ to lower valance states, as highlighted in the inset of Fig. 4(b). The additional small peak on the left side of $L_3$-edge is due to the signal of La $M_4$ from LAO substrates. The Ni *L*-edge XAS spectra of the NNO/CFO heterostructures with thicker CFO layer shows very low signal-to-noise or even vanishes considering the limited probing depth of TEY-XAS. Clearly, the Co and Ni

*L*-edge XAS spectra exhibit opposite trend with the CFO capping layer increasing, which suggests the charge transfer from cobalt to nickel at the NNO/CFO interface. We also characterized the variation of valence state of Fe in the NNO/CFO heterostructures [Fig. 4(c)], which keeps unchanged, as shown in the inset, because Fe ions possess +3 valence in the CFO and it is hard for Fe to loss electrons further. Therefore it is concluded that the charge transfer occurs from Co to Ni ions rather than Fe to Ni.

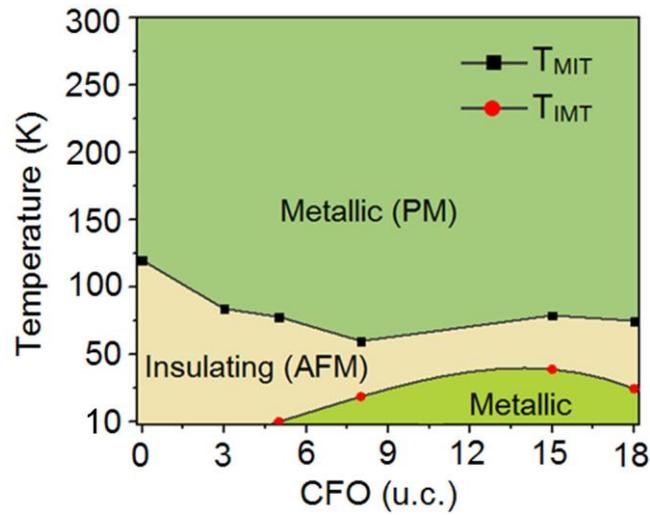

**FIG. 5** Phase diagram painted by $T_{MIT}$ and $T_{IMT}$ data points taken from resistivity measurements. *PM* and *AFM* and *Metallic* region indicate the states of NNO during 300–10 K temperature sweep.

In general, the insulating and antiferromagnetic behaviors emerge simultaneously in NNO,[1,9] which could explain the appearance of the exchange bias behavior and the MIT at around 85 K, as observed in the 3 u.c. sample. In the following, we discuss the origins of the observation of metal-insulator-metal transition and the weaker antiferromagnetic properties at low temperature in thick CFO samples, e.g., NNO (20 u.c.)/CFO (15 u.c.). Nickel possesses one electron in $e_g$ orbit in $Ni^{3+}$ state, as illustrated in Fig. 4(d). Once the charge transfer from Co to Ni occurs, the electron in $e_g$ orbit becomes two with a high spin order, resulting in the enhanced magnetic moments of the nickel. That is, charge transfer at the interface follows $Co^{2+} + Ni^{3+} \rightarrow Co^{3+} + Ni^{2+}$, accompanied by the ferromagnetic interaction between $Ni^{2+}$–$Ni^{2+}$ and $Ni^{2+}$–$Co^{3+}$ ions. The ferromagnetic interactions overcome the antiferromagnetic superexchange of NNO at the interface, providing a double exchange-like mechanism and the resultant weak ferromagnetic ordering in the NNO antiferromagnetic matrix. This is similar to the scenario reported in the previous publication,[8,28–30] and a further low-temperature XMCD (x-ray magnetic circular dichroism) is expected to give an insight into the respective contribution to magnetism during the transition. But this is not the case for the 3 u.c. sample, because only a part of the NNO surface is covered by the CFO particles with respect to the AFM morphologies, which most likely suppresses the macroscopic ferromagnetic coupling and the metallic behavior in a large scale.

Last but not the least, a phase diagram on the basis of our results is depicted in Fig. 5, which summarizes the conductivity behavior of the NNO with the CFO capping layer. Metallic/paramagnetic, insulating/antiferromagnetic and metallic behaviors of NNO at different temperatures are distinguished by different regions with the boundary line. Here the boundary line marked by the extrapolation of $T_{MIT}$ and $T_{IMT}$ data points were taken from transport curves in Fig. 2. The phase diagram reveals that the metallic phase at low temperature can be stabilized by controlling the thickness of the CFO capping layer.

In conclusion, the ultrathin $NdNiO_3$ films with $CoFe_2O_4$ capping (>5 u.c.) show an unusual metallic state at low temperature, which is ascribed to the charge transfer at the interface and the resultant ferromagnetic interaction between Ni–Ni and Ni–Co atoms. The observation of metal-insulator-metal transition in perovskite-spinel oxides might attract extensive investigations on exploring the underlying physics and feasibility of creating electronic devices.

The authors acknowledge Beamline BL08UIA in the Shanghai Synchrotron Radiation Facility for x-ray absorption spectroscopy measurements. This work is supported by the National Natural Science Foundation of China (Grant Nos. 51231004, 51571128, and 51671110) and Ministry of Science and Technology of the People's Republic of China (Grant Nos. 2016YFA0203800 and 2014AA032901).